\documentclass{sig-alternate}
\pdfoutput=1 

\usepackage[T1]{fontenc}
\usepackage[utf8]{inputenc}
\usepackage{microtype}

\usepackage{graphicx}
\usepackage{color}
\usepackage{caption}
\usepackage[numbers]{natbib}
\renewcommand{\cite}[1]{\citep{#1}}

\usepackage{algorithm}
\usepackage{algorithmic}

\usepackage[english]{babel}
\usepackage{booktabs}
\usepackage{xspace}
\usepackage{complexity} 
\usepackage{verbatim} 
\usepackage{enumerate} 
\usepackage{inconsolata} 
\usepackage{flushend}

\usepackage{listings}
\definecolor{mygray}{rgb}{0.5,0.5,0.5}
\definecolor{myblue}{rgb}{0,0,1}
\definecolor{mygreen}{rgb}{0,0.5,0}
\definecolor{myred}{rgb}{0.64,0.08,0.08}
\usepackage[hidelinks]{hyperref}

\newcommand{\pdftitle}{Parameter-Free Probabilistic API Mining across GitHub}

\hypersetup{
  pdftitle={\pdftitle},
  pdfauthor={},
  plainpages=false,
}
\addto\extrasenglish{}
\addto\extrasenglish{}
\addto\extrasenglish{}

\numberwithin{equation}{section} 
\allowdisplaybreaks[3] 
\usepackage{enumitem}
\setitemize{noitemsep,topsep=0pt,parsep=0pt,partopsep=0pt,leftmargin=1.5em}
\setenumerate{noitemsep,topsep=0pt,parsep=0pt,partopsep=0pt,leftmargin=1.5em}

\normalfont
\DeclareFontShape{T1}{aer}{bx}{sc} {<-> ssub * cmr/bx/sc }{}

\newcommand{\cf}{\hbox{{cf.}}\xspace}

\newcommand{\eg}{\hbox{{e.g.}}\xspace}
\newcommand{\ie}{\hbox{{i.e.}}\xspace, } 

\newcommand{\abs}[1]{\lvert #1 \rvert}

\newcommand{\deq}{\mathrel{\mathop:}=}

\renewcommand{\S}{§} 
\newcommand{\calI}[0]{\mathcal{I}}
\newcommand{\zB}[0]{\mathbf{z}}
\newcommand{\piB}[0]{\boldsymbol{\pi}}

\newcommand{\mchoose}[2]{\left(\!\!\left(\begin{array}{@{}c@{}}#1\\#2\end{array}\right)\!\!\right)}
\newcommand{\tmchoose}[2]{\bigl(\!\!\binom{#1}{#2}\!\!\bigr)}
\newcommand{\call}[1]{\texttt{m#1}}
\newcommand{\specialcell}[2][c]{%
  \begin{tabular}[#1]{@{}c@{}}#2\end{tabular}}

\makeatletter
\def\@copyrightspace{\relax}
\makeatother

\newlength{\emstr}
\newcommand{\boldpara}[1]{%
\smallskip%
\par\noindent\textbf{\textit{#1}}\hspace{\emstr}
}%

\newdef{definition}{Definition}

\begin{document} 
%

\title{\pdftitle}

\numberofauthors{2}

\author{%
\alignauthor Jaroslav Fowkes
\and
\alignauthor Charles Sutton
\and%
\end{tabular}\newline\begin{tabular}{c}
\affaddr{School of Informatics} \\
\affaddr{University of Edinburgh, Edinburgh, EH8 9AB, UK} \\
\affaddr{\{jfowkes, csutton\}@ed.ac.uk}
}


\def\@maketitle{\newpage
 \null
 \setbox\@acmtitlebox\vbox{%
\baselineskip 20pt
\vskip 2em                   
   \begin{center}
    {\ttlfnt \@title\par}       
   \end{center}}
 \dimen0=\ht\@acmtitlebox
 \unvbox\@acmtitlebox
 \ifdim\dimen0<0.0pt\relax\vskip-\dimen0\fi}

\maketitle
\begin{abstract} 
Existing API mining algorithms can be difficult to use as they require expensive parameter tuning and the returned set of API calls can be large, highly redundant and difficult to understand. To address this, we present PAM (Probabilistic API Miner), a near parameter-free probabilistic algorithm for mining the most interesting API call patterns. We show that PAM significantly outperforms both MAPO and UPMiner, achieving 69\% test-set precision, at retrieving relevant API call sequences from GitHub. Moreover, we focus on libraries for which the developers have explicitly provided code examples, yielding over 300,000 LOC of hand-written API example code 
from the 967 client projects in the data set.
This evaluation suggests that the hand-written examples actually
have limited coverage of real API usages. 

\end{abstract} 


\section{Introduction}\label{sec:intro}
Learning the application programming interface (API) of an unfamiliar library or software framework
can be a significant obstacle for developers \cite{robillard2009,robillard2011}.
This is only exacerbated by the fact that API documentation can often be
incomplete or ambiguous \cite{uddin2015api}.
Fortunately, an opportunity to address this problem has arisen out of
the simultaneous growth in the amount of source code that is available online
and the growth of large scale data mining and machine learning methods.
This confluence has enabled the development of \emph{API mining} methods \cite{zhong2009mapo,wang2013mining}, which aim to automatically extract a set 
of \emph{API patterns}, which are lists of API methods that are usually used together,
and that together characterize how an API is used.

Despite a number of interesting proposed tools, including well known ones
such as MAPO \cite{zhong2009mapo} and UPMiner \cite{wang2013mining}, so far API
mining tools have not yet gained
 wide-spread adoption in development environments
 such as Eclipse and Visual Studio. 
 We suggest that 
 the fundamental reason for this is that 
 the quality of the extracted patterns
 is not yet high enough: the
 patterns returned by current methods are numerous and highly redundant
 (Section~\ref{sec:numerics}).
For example, \autoref{fig:pam_calls} shows
 the top ten API patterns extracted
 by two state of the art methods.
To a large extent, the patterns from both methods are variations on a theme,
repeating different variations
of ways to repeat the same few API methods.

 The fundamental reason for this, we argue,
 is that current API mining methods
 are built on statistically
 shaky ground.
Specifically, API mining algorithms largely employ \emph{frequent sequence mining}, which is a family of techniques from the data mining literature that takes as input a database of sequences
and from those attempts to identify a set of patterns that frequently occur as subsequences.
In the context of API mining,
each sequence in the database is the
list of API methods called by some client
method, and the subsequence patterns
that are mined are candidates for API patterns.
Frequent sequence mining methods
are very good at their intended purpose,
which is to efficiently enumerate subsequences that occur frequently.
But they are not suitable for pattern mining all by themselves, for a simple reason:
\emph{frequent patterns are often uninteresting patterns.}
This problem is well known in the data mining literature (\cite{aggarwal2014frequent}, Chap.\ 5),
and the technical reason for it is easy to understand. 
Events that are individually frequent but unrelated will also tend to frequently 
occur in a sequence, simply by chance.
Imagine running a frequent sequence mining algorithm on the sequence of events that occur in the day
of a typical academic. Perhaps the most frequent individual events would be
\textsc{SendEmail} and \textsc{DrinkCoffee}.
Then, a frequent sequence miner may well return the pattern
(\textsc{SendEmail}, \textsc{DrinkCoffee}) even if
the two actions have no direct statistical relationship,
because frequent sequence miners do not consider any
notion of statistical correlation or independence 
among items in the sequence.
Within the API mining literature,
methods like MAPO and UPMiner apply clustering
methods precisely to reduce the number of redundant API patterns
that would be returned from the frequent sequence miner;
however, as we show in \autoref{sec:numerics}, even with this step,
substantial redundancy remains.

We address this problem by developing new a mining algorithm
that returns API patterns that not only occur often, but also 
that occur in a sequence more often than expected by chance, 
that is, the \emph{most interesting} sequences (\cite{aggarwal2014frequent}, Chap.\ 5). In order to quantify what ``most interesting'' means, we employ a powerful set of techniques from statistics and machine learning called \emph{probabilistic modelling.}
That is, we design a probability distribution over the list of API methods called
by client methods, and we evaluate a proposed API pattern by whether it improves
the quality of the model. The quality of a probabilistic model can be measured
simply by whether it assigns high probability to a training set.

While numerous sequential probabilistic models of source code have appeared in the literature \cite{hindle2012naturalness,allamanis2014learning,nguyen13statistical,raychev2014code} these all rely on $n$-gram language models.
 But an $n$-gram in this context is a contiguous sequence of method calls, 
 whereas API patterns contain \emph{gaps}, that
 is, the methods in an API pattern do not always occur contiguously in a client method, but instead can have other method calls interspersed between them.
For example, consider the methods in \autoref{fig:ex} which show real-world usage examples of a four line call sequence to set up a twitter client. The first four lines of the last method shows the basic call sequence but the other two methods have additional calls interspersed between them, \eg the first method also turns on SSL support.
Although frequent sequence mining algorithms handle gaps in sequences automatically, previous probabilistic models of code do not,
so we need to introduce a new probabilistic model.

To address this unsatisfactory state of affairs, we propose PAM (Probabilistic API Miner)\footnote{https://github.com/mast-group/api-mining}, a near parameter-free probabilistic algorithm for mining the most interesting API patterns. PAM makes use of a novel probabilistic model of sequences, based on generating a sequence by interleaving a group of subsequences. The list of component subsequences are then the mined API patterns. This is a fully probabilistic formulation of the frequent sequence mining problem, is able to correctly represent both gaps in sequences, and --- 
unlike API mining approaches based on frequent sequence mining --- largely avoids returning
 sequences of items that are individually frequent but uncorrelated. 


Furthermore, another drawback
of current methods for API mining is that
 they have multiple parameters that are very hard to tune in practice, rendering them difficult to use without expensive pre-training.
First, frequent sequence mining methods depend on a user-specified dataset-specific minimum frequency threshold \cite{agrawal1995mining,han2007frequent}. This threshold is extremely difficult to tune in practice as it is prone to exponential blow-up: setting the threshold too low leads to billions of patterns. Conversely, setting it too high leads to no patterns at all. API mining algorithms also tend to employ hierarchical clustering techniques in an attempt to reduce the inherent redundancy of frequent sequence mining (which tends to produce highly repetitive patterns). Choosing the number of clusters/cutoff is considered somewhat of a black art and can only be reliably inferred through computationally expensive training procedures on a validation set \cite{zhong2009mapo,wang2013mining}.
PAM, on the other hand, is near parameter free in the sense that our two user-specified parameters are independent of the dataset, have simple intuitive interpretations and sensible default values.

To showcase PAM, we apply it to mining API patterns for 17 of the most popular Java libraries
 on the GitHub hosting service. We collect all usages of these APIs on GitHub,
yielding a data set of 54,911 client methods from 967 client projects totalling over 4 million lines of code (LOC).
We focus on libraries that contain a specific \texttt{examples/} subdirectory, which allows
us to automatically collect API usage examples that the libraries' developers thought were most salient.
This yields a corpus of 3,385 Java files comprising 300,000 LOC solely of hand-written API usage examples.
These two sources of information allow us to perform extensive evaluation of PAM compared
to MAPO and UPMiner. 
We show that PAM significantly outperforms both MAPO and UPMiner, achieving 69\% precision 
on our test set.
Moreover, the set of patterns that PAM returns is dramatically less redundant than those from MAPO or UPMiner.
To the best of our knowledge, we are the first to mine API calls across the entirety of GitHub, from a data set of 54,911 client methods totalling over 4 million LOC.

We examine whether the API patterns mined by PAM could be used to augment API examples
that are provided with popular projects. This is a new method of evaluating API mining algorithms.
We show that there is substantial overlap between mined patterns and developer-written examples,
providing evidence that the mined patterns are meaningful, but that --- even though our corpus
averages 18,000 example LOC per project --- the mined patterns describe many new patterns
that appear in practice but are not covered by the examples.

\section{Related Work}\label{sec:lit}
The first algorithm for mining API usage patterns from source code was MAPO, proposed by \citet{xie2006mapo} and extended by \citet{zhong2009mapo}. MAPO \cite{zhong2009mapo} mines method call sequences (that call a desired API) from code snippets retrieved by code search engines. Sequences of called API methods are first extracted from the code snippets and then clustered according to a distance metric, computed as an average of the similarity of method names, class names, and the called API methods themselves. For each cluster, MAPO mines the most frequent API calls using SPAM and feeds them to an API usage recommender that ranks them based on their similarity with the developer’s code context. 

UP-Miner \cite{wang2013mining} extends MAPO in an attempt to further reduce the redundancy of the mined API call sequences. This is achieved through three principal approaches: using the BIDE closed frequent sequence miner that returns only the frequent sequences that have no subsequences with the same frequency; using a clustering distance metric based on the set of all API call sequence $n$-grams and an additional clustering step on the mined call sequences from each cluster. Unlike MAPO, the API call sequences are presented as probabilistic graphs, ranked according to their frequency. 

\citet{acharya2007mining} extract API usage scenarios among multiple APIs as partial orders. Frequent API calls are first mined from inter-procedural program traces, API call ordering rules are then extracted from the frequent calls and presented as partial order graphs. 

\citet{buse2012synthesizing} generate API usage documentation by extracting and synthesizing code examples from API method call graphs annotated with control flow information. Their approach first extracts method control flow paths from files that use a desired API. Path predicates are then computed for each statement relevant to a static instantiation of a desired API type and assembled, along with the statements, into API usage graphs. The API usage graphs are then clustered based on their statement ordering and type information, and an abstract usage graph is extracted from each cluster by merging its concrete API usages. Finally, each abstract API usage graph is transformed into a code fragment by extracting the most representative abstract statement ordering from the graph and assigning each abstract statement a concrete name according to  heuristics.

MUSE \cite{moreno2015how} uses static slicing to extract and rank code examples that show how to use a specific method. For each relevant method invocation, MUSE extracts an intra-procedural backward slice, representing a raw code example. The raw code examples are then clustered using type-2 clone detection and the resulting groups of clones ranked according to their popularity. A representative code example from each group of clones is selected based on its readability and ease of reuse, and each example is annotated with inline descriptions of method parameters mined from the Javadoc.

Other API mining papers include \citet{uddin2012temporal} who detect API usage patterns in terms of their time of introduction into client programs and \citet{li2005pr} who automatically extract implicit programming rules in large software code.  
A related approach that has been well-studied in the literature is API code search, where text matching is used to find API usage examples from a large, often online, corpus. SNIFF \cite{chatterjee2009sniff} finds abstract code examples relevant to a natural language query expressing a desired task. SNIFF annotates publicly available source code with API documentation and the annotated code is then indexed for searching. In response to a query, matching code examples are retrieved and concise examples are extracted via a syntax-aware intersection, clustered and ranked based on their frequency of occurrence. Similarly, \citet{keivanloo2014spotting} find abstract code examples by combining textual similarity and clone detection techniques, ranking the returned examples according to their similarity to the query and the completeness and popularity of their concrete usage patterns. \citet{bajracharya2010leveraging} find concrete API usage examples by combining heuristics based on structural and textual aspects of the code, based on the assumption that code containing similar API usages is also functionally similar. 

Strathcona \cite{holmes2005using} is a code example recommendation tool that automatically generates queries from the developer's current code context. The code examples relevant to the queries are identified 
using six heuristics that compare the structural context of the query against the structural context of the classes and methods within an example repository. The resulting code examples are ranked according to their frequency in the final set of top examples returned by each heuristic. Prompter \cite{ponzanelli2014mining} takes a similar approach, matching the generated context-specific queries to Stack Overflow discussions and automatically recommending the discussions most relevant to the developers’ task. 

The API mining problem we consider (Section~\ref{sec:extracting}) is specifically to return sequences
of  API methods that are used together. Other important but distinct data mining problems that are related
to API usage include mining preconditions of API methods \cite{nguyen2014mining}, and mining code changes
\cite{livshits05dynamine,nguyen2013study}.
An important recent piece of infrastructure for large-scale
mining work on code is Boa \cite{dyer2013boa}.
Another interesting line of work is to mine existing code to measure the popularity of language constructs
and APIs \cite{dyer14billions}; our work considers a different mining problem, that of discovering combinations
of API methods that are used together, rather than measuring the popularity of known language features.

Sequential pattern mining is an extremely well studied problem with a huge number of published papers on the topic. The problem was first introduced by \citet{agrawal1995mining} in the context of market basket analysis, which led to a number of other algorithms for frequent sequence mining, including GSP \cite{srikant1996mining}, PrefixSpan \cite{pei2001prefixspan},
 SPADE \cite{zaki2001spade} and SPAM \cite{ayres2002sequential}.
Frequent sequence mining suffers from \emph{pattern explosion}: a huge number of highly redundant frequent sequences are retrieved if the given minimum support threshold is too low.
One way to address this is by mining frequent closed sequences,
\ie those that have no subsequences with the same frequency,
such  as via the BIDE algorithm \cite{wang2004bide}.
More recently, there has been work on sequence mining that directly addresses the pattern explosion issue,
such as SQS-search \cite{tatti2012long}  and GoKrimp algorithm \cite{lam2014mining}.
Our proposed approach falls into this class of probabilistic sequential pattern mining algorithms, and returns patterns that are of a quality that is comparable to, if not better than, both SQS and GoKrimp (see \cite{fowkes2016subsequence} for details).

There have also been sequential probabilistic models of source code proposed in the software engineering literature. \citet{hindle2012naturalness} develop an $n$-gram language model for source code and apply it to a code completion task. \citet{allamanis2014learning} use an $n$-gram language model to learn and suggest variable naming conventions. \citet{raychev2014code} develop an $n$-gram language model that synthesizes code completions for programs using APIs.

\section{Extracting API Calls}\label{sec:extracting}
\begin{figure*}[t]
\begin{lstlisting}
private FinchTwitterFactory(Context context) {
    mContext = context;

    installHttpResponseCache();

    ConfigurationBuilder configurationBuilder = new ConfigurationBuilder();
    configurationBuilder.setOAuthConsumerKey(ConsumerKey.CONSUMER_KEY);
    configurationBuilder.setOAuthConsumerSecret(ConsumerKey.CONSUMER_SECRET);
    configurationBuilder.setUseSSL(true);
    Configuration configuration = configurationBuilder.build();
    mTwitter = new TwitterFactory(configuration).getInstance();
}
\end{lstlisting}
\begin{lstlisting}
public Twitter getTwitterInstance() {
    ConfigurationBuilder cb = new ConfigurationBuilder();
    cb.setOAuthConsumerKey(Keys.consumerKey);
    cb.setOAuthConsumerSecret(Keys.consumerSecret);
    cb.setOAuthAccessToken(mSettings.getString("accessToken", null));
    cb.setOAuthAccessTokenSecret(mSettings.getString("accessSecret", null));
    TwitterFactory tf = new TwitterFactory(cb.build());
    return tf.getInstance();
}
\end{lstlisting}
\begin{lstlisting}
private void startOAuth() {
    ConfigurationBuilder configurationBuilder = new ConfigurationBuilder();
    configurationBuilder.setOAuthConsumerKey(Const.CONSUMER_KEY);
    configurationBuilder.setOAuthConsumerSecret(Const.CONSUMER_SECRET);
    twitter = new TwitterFactory(configurationBuilder.build()).getInstance();

    try {
        requestToken = twitter.getOAuthRequestToken(Const.CALLBACK_URL);
        Toast.makeText(this, "Please authorize this app!", Toast.LENGTH_LONG).show();
        this.startActivity(new Intent(Intent.ACTION_VIEW, 
            Uri.parse(requestToken.getAuthenticationURL() + "&force_login=true")));
    } catch (TwitterException e) {
        e.printStackTrace();
    }
}
\end{lstlisting}
\caption{Three real-world usage examples of a \texttt{twitter4j} API pattern that sets up a twitter client with OAuth authorization.}
\label{fig:ex} 
\end{figure*}

First we describe the specific API mining problem that we consider in this paper.
For every client project that uses a given API, we extract the sequence of API calls used by each method in the project. The problem of \emph{mining API calls} is then to infer from these sequences of API calls those subsequences that represent typical usage scenarios for the API.
These could then be either supplied in example documentation
to client developers, or suggested in real time as developers
type.

For the purposes of this paper, we use a best-effort approach to extract API call sequences directly from source code files. Following MAPO \cite{zhong2009mapo}, we parse each Java source file of interest using the Eclipse JDT parser and extract method calls to API methods from a specified library using a depth-first traversal of the AST. For simplicity, unlike MAPO, we do not attempt to enumerate all possible branches of conditional statements. For example, for the code snippet
\lstinline|if(m1()) { m2(); } else { m3(); }|,
our method returns the call sequence \texttt{m1,m2,m3} whereas MAPO would return the call sequences \texttt{m1,m2} and \texttt{m1,m3}. If this example were indeed
a common API usage pattern, we would argue that returning \texttt{m1,m2,m3} is better in principle, because a subsequence
like \texttt{m1,m2} would provide only incomplete information
about what the developer should write next.

Furthermore, unlike MAPO, we only consider method invocations and class instance creations, and approximately resolve their fully qualified names from the file import statements. For simplicity, superclass methods (and their return types), super method/constructor invocations and class cast expressions are not considered.

We use an approach similar to the original MAPO paper \cite{xie2006mapo} to approximately resolve fully qualified method names. We keep track of field and local variable declarations so that we can resolve the fully qualified name of a method call on the field/variable. We also keep track of import statements so that we can resolve fully qualified names of classes that are explicitly imported, as well as those imported using a wildcard by scanning for specific imports from the wildcarded package in the corpus in a pre-processing step. Additionally, in the pre-processing step we find the return types of locally declared methods so that we are able to subsequently resolve any calls on them. Finally, we filter out any method names that cannot be fully resolved. Each API call sequence is then the sequence of fully qualified API method names that are called by a method in the source file.

For example, consider the client methods in \autoref{fig:ex} that all share the common \texttt{twitter4j} API call sequence:

\begin{lstlisting}
ConfigurationBuilder.<init>
ConfigurationBuilder.setOAuthConsumerKey
ConfigurationBuilder.setOAuthConsumerSecret
ConfigurationBuilder.build
TwitterFactory.<init>
TwitterFactory.getInstance
\end{lstlisting}

This is the minimum API call sequence required to set up OAuth authorization for a twitter client. All the methods in \autoref{fig:ex} have added extra API calls for optional functionality (\eg SSL encryption) but all contain the minimal API call sequence as a subsequence.

There are of course limitations to this approximation (as noted in the original MAPO paper \cite{xie2006mapo}). In particular it is not possible to resolve external nested method calls (\ie in the call \texttt{method1().method2()}, we cannot resolve \texttt{method2()} unless \texttt{method1()} is declared locally). However, for the purposes of this paper we are primarily interested in assessing the performance of PAM.
Moreover, it is important to note that PAM is flexible and supports any API call extractor that returns sequences of calls, making it applicable to dynamically inferred call sequences as well as other programming languages. While we mine (possibly incomplete) API calls that are inferred statically from \texttt{.java} files in this paper, one can in principle extract fully resolved static or dynamic API call sequences using the BCEL bytecode library \cite{bcel,java-callgraph}. The reason we did not perform dynamic call sequence extraction is that the idiosyncratic build process of most Java projects would have made compiling all $967$ open-source Java projects that used our chosen libraries in our dataset (see \autoref{tab:examples}) prohibitive. Finally, note that any API call extractor that only extracts sequences of calls cannot handle conditional statements properly, as it is trying to approximate a graph with a sequence.

\section{Mining API Call Patterns}\label{sec:mining}
In this section we will describe our novel
probabilistic model for API mining.
Our model is a joint probability distribution
over the list of API calls in a client method,
which we observe in the data, and the underlying
API patterns that the programmer intended to use,
which we never observe directly.
The model defines this probability assuming that the set of all possible true
API patterns is known. Then, learning involves working backward: given the client
methods that were observed, what set of true API patterns might have generated them?
Specifically, we measure the quality of a proposed set of API patterns
by supposing those were the true patterns, and measure
the probability that the model assigns to all client methods in the
database. We 
search for the set of API patterns that maximizes
this probability, specifically we perform this search under the framework
of a celebrated algorithm from statistics called expectation-maximization (EM) \cite{dempster1977maximum}, which has seen an enormous
number of applications.
We use a particular variant called
\emph{structural EM} \cite{friedman1998bayesian},
as this deals with the specific setting of
learning via search through a large combinatorial space,
 in our case, the space of all possible sets
of API patterns.

In the next sections, we give a high-level overview of each
of the aspects of the model. For ease of exposition, 
we do not describe some of the more theoretical aspects; for those we refer the reader to our paper \cite{fowkes2016subsequence} that describes a similar model and algorithm for general sequential pattern mining. 

\subsection{Probabilistic Model}
\label{sec:model}

In this section, we describe the probabilistic model that PAM is based on.
The model is a probability distribution that, based on a set of API patterns, defines a distribution over all possible API patterns present in client code. 
When a probabilistic model becomes more complex than
one of the well-known standard families, then it is often easiest to explain by describing an algorithm to sample from it. This is how we will proceed in this section.

The model has two different types of parameters: a set of API patterns and a set of probabilities.  The API patterns 
are straightforward: each API pattern is a sequence $S_a = (a_1, \ldots a_n)$ of method names from the API. 
We allow patterns to occur more than once in the same client method.
Therefore, for each API pattern $S_a,$ the model also includes a probability distribution over the integers $0,1,\ldots M$ which represents how likely a client method is to include the pattern 
$S$ zero times, one time, etc.
We define $\calI$ to be the set of all API patterns $S_a$ in the model.
We assume that $\calI$ also contains singleton sequences
$(m)$ for every method $m$ in the API --- although an API pattern with only one method call is not very useful, so we never return such patterns
to the user, we will see in \autoref{sec:inference} that
these are a technical device that is necessary for the inference procedure.

Now we present an algorithm that will draw samples from our model,
which we call the \emph{generative algorithm}.
The generative algorithm says:
hypothetically speaking, if our model were correct, how would each client method
be generated assuming that the API patterns
and probabilities are known?
We emphasize that the generative algorithm is simply an explanatory tool
that helps in understanding our approach, and  
is never used while performing the API mining.
The algorithm has two main phases: First, 
from the set of all interesting API patterns,
we sample which ones will appear in the client method
that we are about to generate, and how many times they will be used, which yields a multiset
that we call $\mathcal{S}$.  Then we randomly sample
a way to interleave the sampled API patterns,
and this results in a hypothetical
client method. More formally:
\begin{enumerate}
\item For each unique API pattern $S$ in the set of interesting API patterns $\calI$, decide independently the number of times $S$ should be included in the client API sequence $X$, \ie draw the count $c_S$ from a suitable probability distribution over the integers. 
 \item Set $\mathcal{S}$ to be the multiset with counts $c_S$ of all the API patterns $S$ selected for inclusion in $X$, that is, $\mathcal{S} \deq \{ S \colon c_S \ge 1 \}.$
As $\mathcal{S}$ is a multiset, a single API pattern can occur more than once in $\mathcal{S}$.
 \item Set $\mathcal{P}$ to be the set of all possible sequences that can be generated by interleaving together the API patterns in the multiset $\mathcal{S}$, \ie
 \[\mathcal{P} \deq \{X \colon \,\mathcal{S} \text{ partition of } X, S \subset X \; \forall  S \in \mathcal{S}\},\]
 (see the discussion below for an illustrative example).
 \item Sample $X$ uniformly at random from $\mathcal{P}$.
\end{enumerate}
This algorithm defines a probability distribution over client methods,
which we can sample from simply by executing it.
First, let us clarify the meaning of the set $\mathcal{P}$ through
an example.  To interleave two sequences $S_1$
and $S_2$, we mean the placing of items from $S_1$ into the gaps between items in $S_2$.
For example, if $S_1 = (\call{1},\call{2})$ and
 $S_2 = (\call{3},\call{4})$, then the set 
 of all ways to interleave $S_1$ and $S_2$ is
 \begin{align*}
\mathcal{P} = \{
&(\call{3},\call{4},\call{1}, \call{2}), 
  (\call{3}, \call{1}, \call{4}, \call{2}), \\
&  (\call{3}, \call{1}, \call{2}, \call{4}), 
  (\call{1}, \call{3}, \call{4}, \call{2}), \\
&  (\call{1}, \call{3}, \call{2}, \call{4}), 
  (\call{1}, \call{2}, \call{3}, \call{4})\}.
\end{align*}
It is possible to uniformly
sample from $\mathcal{P}$ efficiently by merging 
in subsequences one at time, but we omit the details 
as it is unnecessary in practice.
 
At this point, the generative algorithm
may seem contrived. Certainly we hope
that developers do not write code in a manner that is anything like this.
To assuage the reader's conscience, we point out that
popular methods such as the $n$-gram language model and
latent Dirichlet allocation, which have been widely applied
both to natural language and programming language text,
also have generative algorithms, and those algorithms
are similarly contrived.  The reason that these models
are useful anyway is that we are primarily interested
not in the \emph{forward} generative direction,
in which we use API patterns to generate client methods,
but in the \emph{reverse} direction, in which we run the generative
algorithm backward to use client methods to infer API patterns.
As we will see in a moment, the backward version of this
model is much more intuitive and natural.


We have so far defined a probability distribution implicitly using a
generative algorithm, however we can now define it explicitly, by giving
a formula for the probability of a client method $X$ under our model.
To do this, we need to introduce notation to handle the fact that our
model allows the same API pattern to occur multiple times in a single
client method. We will consider each occurrence of an API pattern $S$ in a client API sequence $X$ separately: let $S^{[n]}$ denote the $n$-th occurrence of $S$ in $X$ \ie by the notation $(\call{1}, \call{2})^{[3]}$ we mean ``the 3rd time the API pattern $(\call{1}, \call{2})$ occurs in a client sequence''. For example, $X = (\call{1},\call{2},\call{3},\call{1},\call{2})$ contains the API patterns $(\call{1},\call{2})^{[1]}$ and $(\call{1},\call{2})^{[2]}$ \ie the first and second occurrences of $(\call{1},\call{2})$. 

In light of this $p( (\call{1}, \call{2})^{[3]} \in X)$ is naturally defined as the ``probability of seeing $(\call{1}, \call{2})$ for the 3rd time given that we've seen it for the 2nd time'', \ie $p( (\call{1},\call{2})^{[3]} \in X | (\call{1},\call{2})^{[2]} \in X) = p( (\call{1},\call{2})^{[3]} \in X | (\call{1},\call{2},\call{1},\call{2}) \in X)$ (since we are allowing gaps in API patterns and so $(\call{1},\call{2})^{[2]} \in X = (\call{1},\call{2},\call{1},\call{2}) \in X$). Formally, the associated probability $\pi_{S^{[n]}}$ for the $n$-th occurrence $S^{[n]}$ is simply the conditional probability of seeing $S^{[n]}$ in a client sequence $X$ given the previous occurrence $S^{[n-\call{1}]}$, \ie $\pi_{S^{[n]}} = p(S^{[n]})/p(S^{[n-1]})$. 

We also introduce the binary variable $z_{S^{[n]}} \deq \mathbf{1}_{\{c_S \ge n\}}$ to indicate whether $S^{[n]}$ is included in $X$ or not. For clarity of exposition we will drop the explicit occurrence superscript in the sequel.

Now we can give an explicit formula for the probability of a client method $X$
under our model.
Given a set of informative API patterns $\calI$, let $\zB, \piB$ denote the vectors of $z_S,\pi_S$ for all API patterns $S \in \calI$. Assuming
$\zB, \piB$ are fully determined, 
the generative model implies that the probability of generating a client API sequence $X$ is:
\begin{equation*}
 p(X, \zB | \piB) =
\begin{cases} \frac{1}{\abs{\mathcal{P}}}
\prod_{S \in \mathcal{I}}
\pi_S^{z_S}(1-\pi_S)^{1-z_S} \! &\text{if $X \in \mathcal{P}$,} \\
\,0 \! &\text{otherwise}.
\end{cases}
\end{equation*}
Calculating $\abs{\mathcal{P}}$ may seem problematic, however it turns out to be rather straightforward. Pick an arbitrary ordering $s=1,\dotsc,\abs{\mathcal{S}}$ for the selected API pattern $S \in \mathcal{S}$ and observe that in our iterative sampling algorithm, when merging $S_a$ into $S_b$, we have $\abs{S_b}+1$ points to splice $\abs{S_a}$ elements into, that is $\abs{S_b}+1$ multichoose $\abs{S_a}$, denoted $\tmchoose{\abs{S_b}+1}{\abs{S_a}} = \binom{\abs{S_b}+\abs{S_a}}{\abs{S_b}}$.
To see how to compute this, consider a sequence of $k = |S_a|$ stars ($\ast$) and $n = |S_b|$ bars ($|$). The number of
ways that these two sequences can be spliced together is equal to the number of ways to place the $k$ stars into $n+1$ bins delimited by the bars (this is by definition $n+1$ multichoose $k$). This can be equivalently viewed as the number of ways to arrange the $n$ bars amongst the $k$ stars and $n$ bars (which is clearly $n+k$ choose $n$). Applying this formula iteratively we obtain:
\begin{equation*}
\abs{\mathcal{P}} = \prod_{s=1}^{\abs{\mathcal{S}}} \mchoose{1 + \sum_{t=1}^{s-1}\abs{S_t}}{\abs{S_s}}
= \prod_{s=1}^{\abs{\mathcal{S}}} \dfrac{\left(\sum_{t=1}^{s}\abs{S_t}\right)!}
{\abs{S_s}!\left(\sum_{t=1}^{s-1}\abs{S_t}\right)!}.
\end{equation*}


\subsection{Inference}
\label{sec:inference}

\emph{Inference} in a probabilistic model is the task of running a generative procedure backwards. In our model, this is the task of: given a client method $X$, infer the vector $\zB$, which indicates which API patterns
were used in the method.  While the generative algorithm provides
a way to sample from the joint distribution $p(X, \zB|\piB)$,
inference is focused on the conditional distribution $p(\zB|X, \piB).$  Essentially the inference procedure amounts to computing a partition of
the API calls in $X$ according to the mined API patterns.
 
At this point the reader may well be wondering why we need to do inference at all? For a specific client method
$X$, can't we just look up the API patterns that $X$ subsumes? The answer is 
that the mined API patterns overlap, and this is what we want,
because we would like to be able to learn more general and more specific versions of the same pattern.
Consider the case where we have to choose between returning a more general pattern such as:
\begin{lstlisting}
builder.<init>
builder.setCommonProperty
builder.build
\end{lstlisting}
and a more specific one such as:
\begin{lstlisting}
builder.<init>
builder.setCommonProperty
builder.setRareValue
builder.build
\end{lstlisting}
We want these two patterns to compete with each other to explain
each client method, and the fact that we use a partitioning 
approach means that each API call in a client can be explained by at most
one API pattern.  At the end of the day, the effect is that the more specific 
pattern will only survive into the final list of mined patterns 
if it manages to be used to explain enough client methods.
 In other words, because we have \emph{probabilities} associated with each of the two API patterns, we are able to choose the \emph{more interesting} of the two in a well-defined and rigorous statistical way.

More formally, the inference procedure assumes that the vector of probabilities $\piB$ is known (we learn it in the next section). 
To infer the best $\zB$ for a client API sequence $X$, 
the natural probabilistic way of asking this question
is using the conditional distribution $p(\zB|X,\piB).$
Specifically, we search for the vector $\zB$
that maximizes $\log p(\zB|X,\piB)$.
Sadly, it can be shown that this problem is NP-hard in general.
Happily, this problem can be approximately solved using a simple greedy algorithm (\cf Algorithm 1 in \cite{fowkes2016subsequence}),
and we find that the greedy algorithm works well in practice. 
The greedy algorithm repeatedly chooses an API pattern $S$ that maximizes the
improvement in log probability divided by the number of methods in $S$ 
that have not yet been explained.
In order to minimize CPU time,
we cache the
API patterns and coverings for each API client sequence as needed.

Now we can see why
we have insisted on including singleton sequences for API patterns into
$\calI$ even though we would never return them to a user.
Including the singleton sequences, and allowing them to be
repeated arbitrarily many times, ensures that every client
method has at least one valid partitioning.

\subsection{Learning}
\label{sec:learning}

Given a set of interesting API patterns $\calI$, consider now the case where both variables $\zB, \piB$ in the model are unknown. 
There is an obvious chicken and egg problem here: the most interesting API patterns $S$ are determined by maximizing
$\log p(\zB|X,\piB)$ for $z_S$, which
means we we need to know $\pi_S$. 
But the probability $\pi_S$ of an API pattern $S$ depends on how often that pattern is used in a client API sequence $X$. 
To get round this we can use the expectation maximization (EM) algorithm
\citep{dempster1977maximum} which is an algorithm for estimating parameters in a model that has unobserved variables.
 The EM algorithm is a clever way to get around the
 chicken and egg problem, by iteratively solving for
 the best $\zB$ given the current value 
 of $\piB$, then solving for the best value
 of $\piB$. Of course, EM needs an initial guess for the unobserved variables $\piB$ and a good first guess is simply the relative support (\ie relative frequency of occurrence) of each API pattern in $\calI$ (in fact this guess is correct if all the API patterns in $\calI$ are independent of each other). For the mathematical details, we refer the interested reader to Algorithm 2 in \cite{fowkes2016subsequence}.

\subsection{Inferring new API patterns}
Now that we have shown how to learn the parameters of our probabilistic model, the astute reader may well note that we still haven't got anywhere as we have no way of inferring which sequences to include in the set of interesting API patterns $\calI$. However, we can once again turn to the statistical community for a solution, in the form of the \emph{structural} EM algorithm \citep{friedman1998bayesian}. As the name suggests, this is a variant of the EM algorithm that lets us grow the set of interesting API patterns $\calI$. In particular, we can add a candidate API pattern $S'$ to the set of interesting API patterns $\calI$ if doing so
improves the value of $\log p(\zB|X,\piB)$ averaged
across all client API sequences $X$.

To get an estimate of maximum benefit to including
candidate $S'$, we must carefully choose an initial value of
$\boldsymbol\pi_{S'}$ that is not too low, to avoid getting stuck in a local optimum.
To infer a good $\boldsymbol\pi_{S'}$, we force the candidate $S'$ to explain all
API client sequences it supports by initializing $\boldsymbol\pi_{S'} = (0,1,\dotsc,1)^T$ and update $\boldsymbol\pi_{S'}$
with the probability corresponding to its actual usage once we have inferred
all the $z_S$. As before, structural EM also needs an initial guess for the set of interesting API calls $\calI$ and associated probabilities $\piB$ and we can simply initialize $\calI$ with all the API methods in the dataset and $\piB$ with their relative supports. Once again, we omit the mathematical details of the algorithm, but refer the interested reader to Algorithm 3 in \cite{fowkes2016subsequence}.


\subsection{Candidate Generation}
However, we are not quite done yet. The structural EM algorithm requires a method to generate new candidate sequences $S'$ that are to be considered for inclusion in the set of interesting API patterns $\calI$. One possibility would be to use a standard frequent sequence mining algorithm to recursively suggest larger API patterns starting from all the API methods in the dataset,
however preliminary experiments found this was not the most efficient method. 
For this reason we take a somewhat different approach and recursively
combine the interesting API patterns in $\calI$ with the \emph{highest
support first}. In this way our candidate generation
algorithm is more likely to propose viable candidate API patterns earlier and  we find that this heuristic works well in practice. Although the algorithm is straightforward, it adds little to the exposition and we omit the details here and refer the interested reader to Algorithm 4 in \cite{fowkes2016subsequence}.

\subsection{Mining Interesting API Patterns}\label{sec:ism}
Finally, we can present our complete Probabilistic API Mining (PAM) algorithm in \autoref{alg:ism}.
\begin{algorithm}[!h]
\caption{Probabilistic API Miner (\textsc{PAM})}\label{alg:ism}
\algsetup{indent=2em}
\begin{algorithmic}
\REQUIRE Client method API call sequences $X^{(1)},\dotsc,X^{(m)}$
\STATE Initialize $\calI$ with singleton API patterns and $\piB$ with their
supports
\WHILE{not converged}
\STATE Add API patterns to $\calI, \piB$ using structural EM \\ \hspace{2em}(Algorithm 2 in \cite{fowkes2016subsequence})
\STATE Optimize parameters for $\calI, \piB$ using EM
\\ \hspace{2em}(Algorithm 1 in \cite{fowkes2016subsequence})
\ENDWHILE
\STATE Remove all singleton API patterns from $\calI$
\RETURN $\calI,\piB$
\end{algorithmic}
\end{algorithm}
As all operations on API client sequences in our algorithm are independent, and
so trivially parallelizable, we perform the $E$ and $M$-steps in both
the EM and structural EM algorithms in parallel.
  
We can rank the retrieved API patterns according to their \emph{interestingness}, that is how likely they are under the probabilistic model, and therefore we rank the API patterns $S\in \calI$ according to their probabilities $\pi_S$ under the model. 

An important property of formulating PAM as a pattern covering problem on each API client sequence is that it strongly favours adding only API patterns of associated methods, \ie methods that largely co-occur in the code. 

\section{Experiments}\label{sec:numerics}
\begin{table*}
\vspace*{8pt}
  \caption{\textsc{Example} dataset extracted from the GitHub Java corpus. Each row is a separate library or framework
  for which we mine a set of API patterns.  Each \emph{Client file set} contains all source files that import a class belonging to the respective package or one of its subpackages. Each \emph{Example file set} contains all source files that are present in the project's example directory. Note that both file sets exclude duplicate files.}
  \small\centering
  \begin{tabular}{llrrl}
  \toprule
  Project & Package Name & Client LOC & Example LOC & Description \\
  \midrule
AndEngine & \texttt{org.andengine} & 18,274 & 19,529 & Android 2D OpenGL game engine \\
Apache Camel & \texttt{org.apache.camel} & 141,454 & 15,256 & Enterprise application integration framework \\
Cloud9 & \texttt{edu.umd.cloud9} & 35,523 & 10,466 & Cloud-based IDE \\
Drools & \texttt{org.drools} & 187,809 & 15,390 &  Business rules management system \\
Apache Hadoop & \texttt{org.apache.hadoop} & 1,951,653 & 26,162 & Map-reduce framework \\
HornetQ & \texttt{org.hornetq} & 30,564 & 22,541 & Embeddable asynchronous messaging system \\
Apache Mahout & \texttt{org.apache.mahout} & 48,206 & 11,772 & Scalable machine learning environment \\
Neo4j & \texttt{org.neo4j} & 239,825 & 7,710 & Graph Database \\
Netty & \texttt{io.netty}& 8,196 & 9,725 & Network application framework \\
RESTEasy & \texttt{org.jboss.resteasy} & 131,447 & 16,055 & RESTful application framework \\
Restlet Framework & \texttt{org.restlet} & 208,395 & 41,078 & RESTful web API framework \\
Spring Data MongoDB & \specialcell[t]{\texttt{org.springframework}\\\texttt{.data.mongodb}}& 16,567 & 18,786 & Spring framework MongoDB integration \\
Spring Data Neo4J & \specialcell[t]{\texttt{org.springframework}\\\texttt{.data.neo4j}} & 6,600 & 9,915 & Spring framework Neo4j integration \\
Twitter4J & \texttt{twitter4j} & 96,010 & 6,560 & Twitter API \\
Project Wonder & \texttt{com.webobjects} & 375,064 & 37,181 & WebObjects frameworks \\
Weld & \texttt{org.jboss.weld} & 23,298 & 9,489 & Contexts and Dependency Injection API \\
Apache Wicket & \texttt{org.apache.wicket} & 564,418 & 33,025 & Web application framework \\[0.5ex]
\midrule
  TOTAL & & 4,083,303 & 310,640 \\
  \bottomrule
  \end{tabular}
  \label{tab:examples}
\end{table*}

In this section we perform a comprehensive evaluation of PAM across GitHub, comparing and contrasting it against MAPO and UPMiner. In particular, we aim to answer the following three research questions.

\boldpara{RQ1: Are the API patterns mined by PAM more prevalent?}
\textmd{This research question evaluates the quality of the API call sequences mined by PAM, as we would expect a set of more representative API call sequences to be more prevalent in a held-out corpus of code. 
By performing a random 50/50 split of a suitable API call dataset, we can see if sequences mined from one half of the dataset are prevalent on the other half and thus if they are representative. Note that performing such a test/train split is standard practice in the evaluation of machine learning algorithms.}
\boldpara{RQ2: Are the API patterns mined by PAM more diverse?}
\textmd{We would also expect a more representative set of API patterns to have lower redundancy, 
as a list in which every pattern uses the same few methods will be both redundant and non-diverse. However a redundancy of zero is not necessarily desirable: As mentioned previously (\autoref{sec:inference}),
a good list of patterns may contain both more general and more specific versions of the same pattern.
That said, a highly redundant list is clearly problematic.}

\boldpara{RQ3: Could the API patterns mined by PAM supplement existing developer-written API examples?}
\textmd{Finally, we investigate if the mined API patterns can be useful in practice. To do so, we look at libraries
and frameworks that explicitly contain directories of API examples provided by the library's developers. This allows us to measure whether API call sequences present in API example directories are returned by PAM, and also vice versa, \ie whether the hand-built example directories
can be improved by including API patterns mined from client code by PAM. We will show both that (a) there is substantial
overlap between the mined patterns and the developer-written
examples, indicating that PAM does indeed find patterns
that the project developers believe are meaningful, but also (b)
PAM identifies a large number of patterns that do not
occur in examples, which could serve as a useful supplement.}

\boldpara{Evaluation Metrics}
A good measure of the quality of mined API call sequences is to see what proportion are contained in a set of relevant \emph{gold standard} sequences, a measure we term \emph{sequence precision}. This allows us to measure the degree to which the mined patterns represent relevant API call sequences. Similarly, we also define \emph{sequence recall} as the proportion of relevant gold standard sequences that contain a mined call sequence. This allows us to measure the degree to which the API miner is able to retrieve relevant API call sequences. 
In other words, sequence precision measures the percentage of mined sequences that are somewhere used, and 
sequence recall measures the degree to which the mined sequences cover the usages in the gold standard
data set. 
We present these metrics in a \emph{precision/recall curve}, as is standard practice in the information retrieval
literature \cite{manning:irbook}. Each point on the precision/recall curve corresponds to a different point in the
ranked list of API patterns returned by each method, and indicates what the sequence precision and sequence recall would be 
if we forced the method to stop returning patterns at that point in the list.  In a precision/recall curve, being up and to the right is better,
as it means that the system returns more accurate results for any fixed recall value. For MAPO and UPMiner, we rank
the list of API patterns by support, whereas for PAM we rank the patterns by their probability under PAM's statistical model.
As for redundancy, we  measure how redundant the set of mined sequences is by calculating the average over each API pattern of the number of other, larger API patterns that contain it (we call this \emph{no.\ containing sequences}).

\boldpara{Dataset}
In order to asses the performance of PAM and perform a thorough comparison with MAPO and UPMiner we assemble a data set of target libraries and frameworks from the GitHub Java corpus \cite{githubCorpus2013}.
We focus on those projects that contain an \texttt{examples/} directory of code examples, so that we can compare
mined patterns to those written by the library's developers.
We include in our data set all Java projects on Github that are (1) sufficiently popular, (2) imported by a sufficient number
of other projects, and (3) that contain a sufficiently large \texttt{examples/} directory.

Specifically, we first find all Java projects in the corpus that have an example directory (\ie matching \lstinline+*example*|*Example*+) containing more than 10K LOC. From these projects we then select those that are in the top $1,000$ projects in the corpus, ranked according to popularity. Popularity in the GitHub corpus is
calculated as the sum of the number of project forks and watchers, where each is separately normalized into a z-score. From these top projects, we determine which of these are called from 50 or more methods belonging to other projects in the corpus, leaving us with the 17 projects in \autoref{tab:examples}.  We call this set of projects and associated client code the \textsc{Example} dataset, 
to emphasize the fact that we focus on libraries and frameworks that include examples.

Each of these 17 projects is a library or framework, which we will call
a \emph{target project}, for which we wish to extract API patterns.
For each target project, we perform API mining separately, and all results are reported as an average
over the 17 target projects.
To extract a set of client methods for each target project, we search the entire GitHub Java corpus for all source files that import a class belonging to the respective package or one of its subpackages and this set of files (excluding duplicates) formed the \emph{Client file set}. Extracting, for each project, all source files in the aforementioned example directory (excluding duplicates) formed the \emph{Example file set}. Statistics on both file sets are given in \autoref{tab:examples}.   

\begin{figure}[tb]
 \centering
 \includegraphics[scale=0.39]{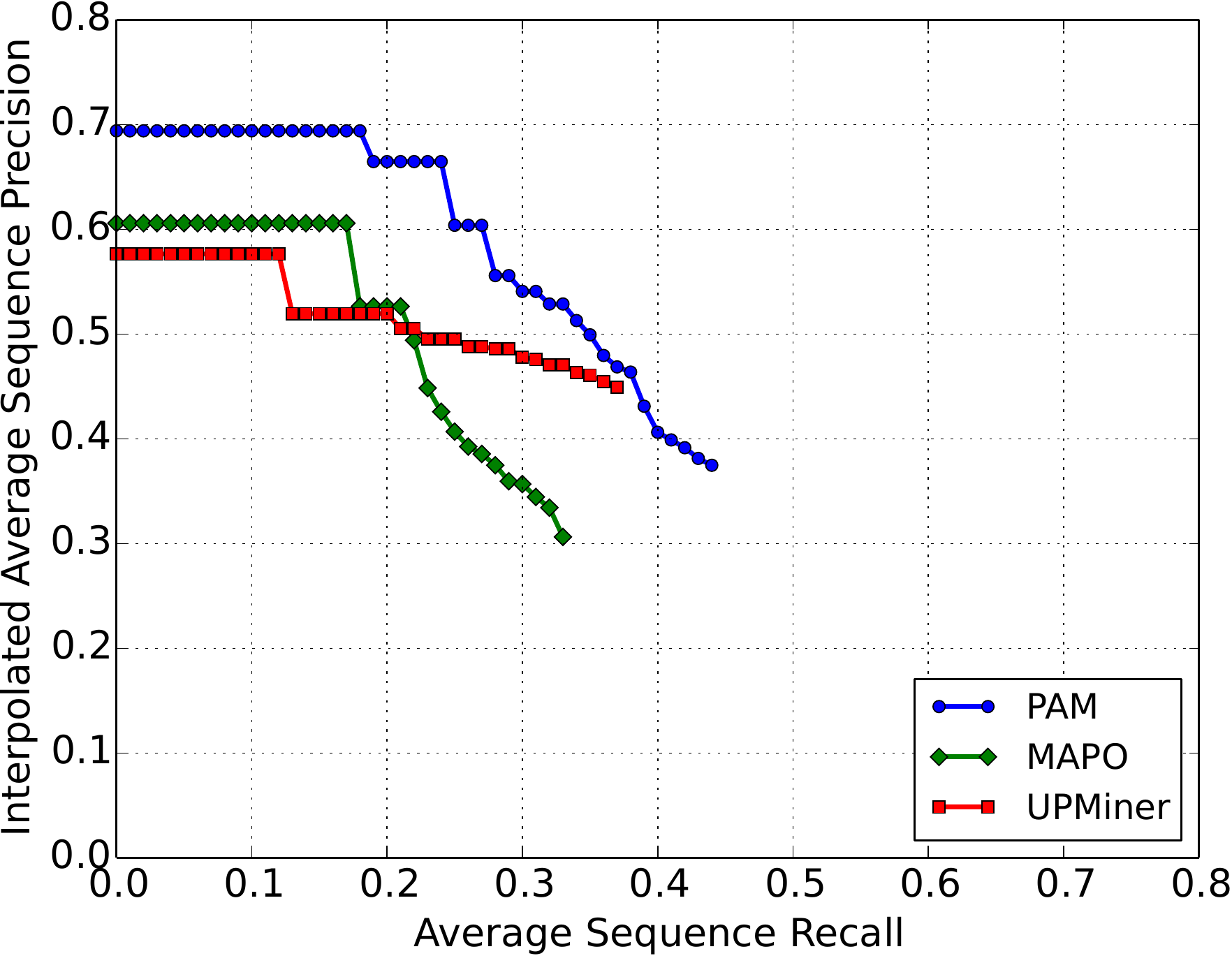}
 \caption{Average \emph{test-set} precision against recall for PAM, MAPO and UPMiner on the \textsc{Example} dataset, using the top-$k$ mined \emph{train-set} sequences as a threshold. 
}
\label{fig:train}
\vspace{-11pt}
\end{figure}


\boldpara{Experimental Setup}
As public implementations were unavailable, we implemented MAPO \cite{zhong2009mapo} and UPMiner \cite{wang2013mining} based on the descriptions in their respective papers. We used a clustering threshold of $40\%$ for MAPO as this gave consistent performance and $20\%$ for UPMiner as this matched the natural cutoff in the dendrogram. The minimum support thresholds for both algorithms were set as low as was practically feasible for each run. We ran PAM for $10,000$ iterations with a priority queue size limit of $100,000$ candidates.

\boldpara{RQ1: Are the API call sequences mined by PAM more prevalent?}
As previously mentioned, in an attempt to answer this question we divide our dataset of API calls in half and see if sequences mined from one half of the dataset are prevalent in the other half. Specifically, we randomly divide the Client file set (\cf \autoref{tab:examples}) into two (roughly) equal train and test sets.
This enables us to mine API call subsequences from the training set and evaluate them using the sequence precision and recall metrics against the API call sequences in the test set. \autoref{fig:train} shows the sequence precision against recall, averaged across all projects in the dataset. 
It is evident that PAM has significantly higher precision and recall than both MAPO and UPMiner, reaching a precision of $69\%$. MAPO performs especially poorly, as its precision degrades significantly as the recall increases. We can therefore say with certainty that the API call sequences mined by PAM are more prevalent. Note that while the best recall that PAM achieves is $44\%$, this is actually close to the theoretical maximum recall on the test set. This can be approximated by the proportion of test set sequences that contain training set sequences, which is around $45\%$. 


\begin{figure}
 \centering
 \includegraphics[scale=0.39]{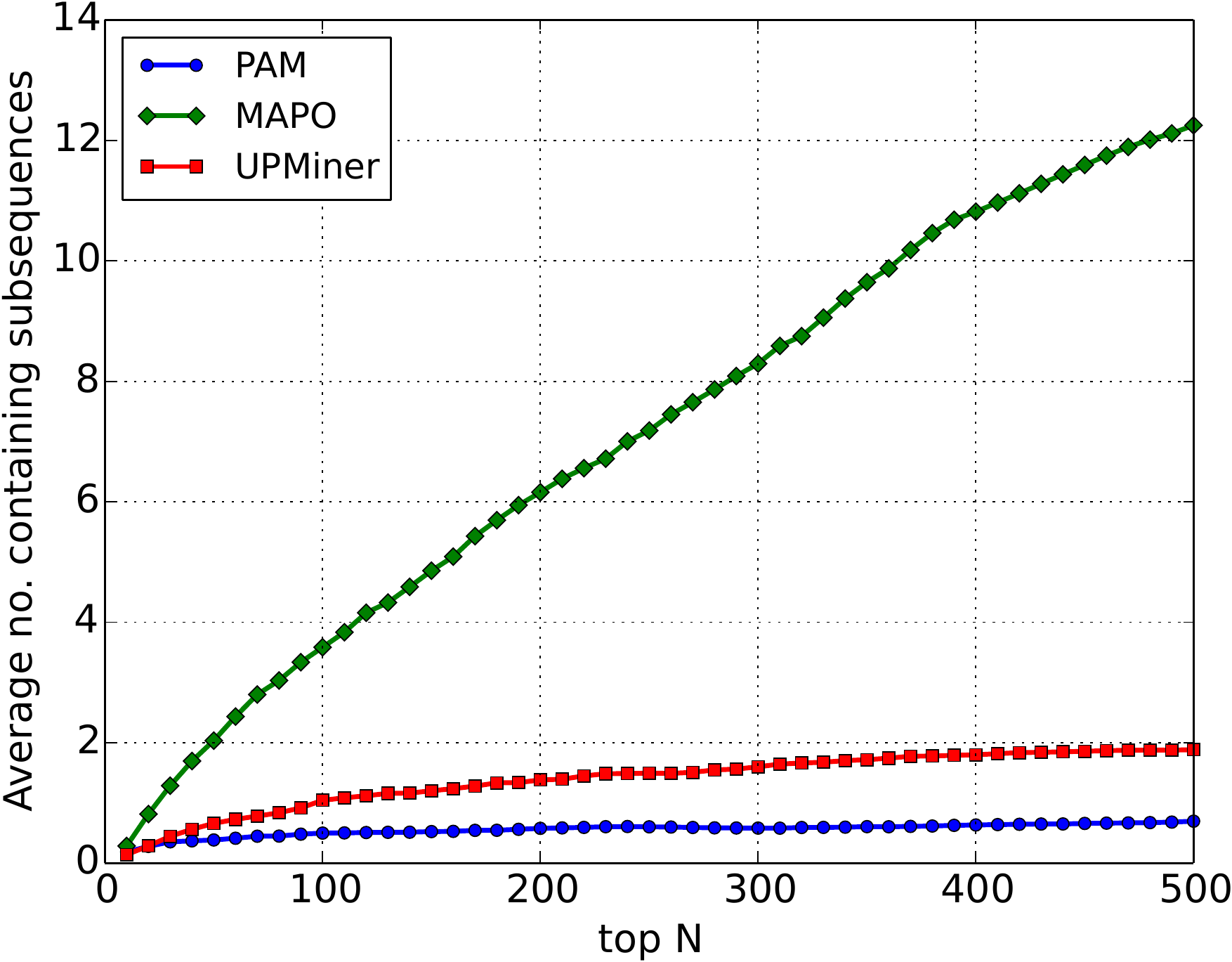}
 \caption{Average no.\ containing sequences for PAM, MAPO and UPMiner on the \textsc{Example} dataset, using the top-$k$ mined sequences.}
\label{fig:all_cs}
\vspace{-11pt}
\end{figure}

\boldpara{RQ2: Are the API call sequences mined by PAM more diverse?}
We now turn our attention to the complete dataset and mine call sequences from the entire Client file set for each project (\cf \autoref{tab:examples}). We can then use the no.\ containing sequences metric to determine how redundant the set of mined call sequences is. \autoref{fig:all_cs} shows the average no.\ of sequences containing other sequences in the set of top-$k$ mined sequences as $k$ varies. One can see that PAM has consistently the lowest figure, around $0.5$, showing that it is the least redundant and therefore most diverse.
One of the key motivations of our method is that the list
of patterns returned by sequence mining methods is redundant.
This figure shows that, even after the extra steps
that MAPO and UPMiner take to reduce the redundancy
of the raw output of frequent sequence mining, 
the patterns returned by PAM are still less redundant.

\boldpara{RQ3: Could the API patterns mined by PAM supplement existing developer-written API examples?} 
We measure whether the mined API patterns correspond to hand-written examples
in the dataset. We therefore mine, for each project, call sequences from the Client file set and evaluate them against call sequences in the Example file set. \autoref{fig:all} shows the sequence precision against recall, averaged across all projects. Again, PAM has evidently better precision and recall than MAPO and UPMiner. The best recall achieved by PAM is $28\%$,
significantly better than the other methods, and for any fixed recall value, PAM has higher precision than the other methods.
This suggests that the API patterns returned by PAM could better supplement developer-written examples than those returned by MAPO or UPMiner.

In an absolute sense, the level of agreement between PAM and the hand-written examples,
although substantial, might not seem especially high. This raises an interesting question: Does this level
of disagreement occur because
the PAM patterns are not representative of the client code they were mined from, \emph{or} because the hand-written examples themselves are not fully representative of the client code?
Although previous work has explored what it means for a \emph{single} API example to be useful \cite{ying2014,seyed2012},
there seems to be much less work about what it means for a \emph{set of examples} to be useful,
and how well example directories in popular projects reflect 
all of the actual common uses of an API. 

\begin{figure}[t]
 \centering
 \includegraphics[scale=0.39]{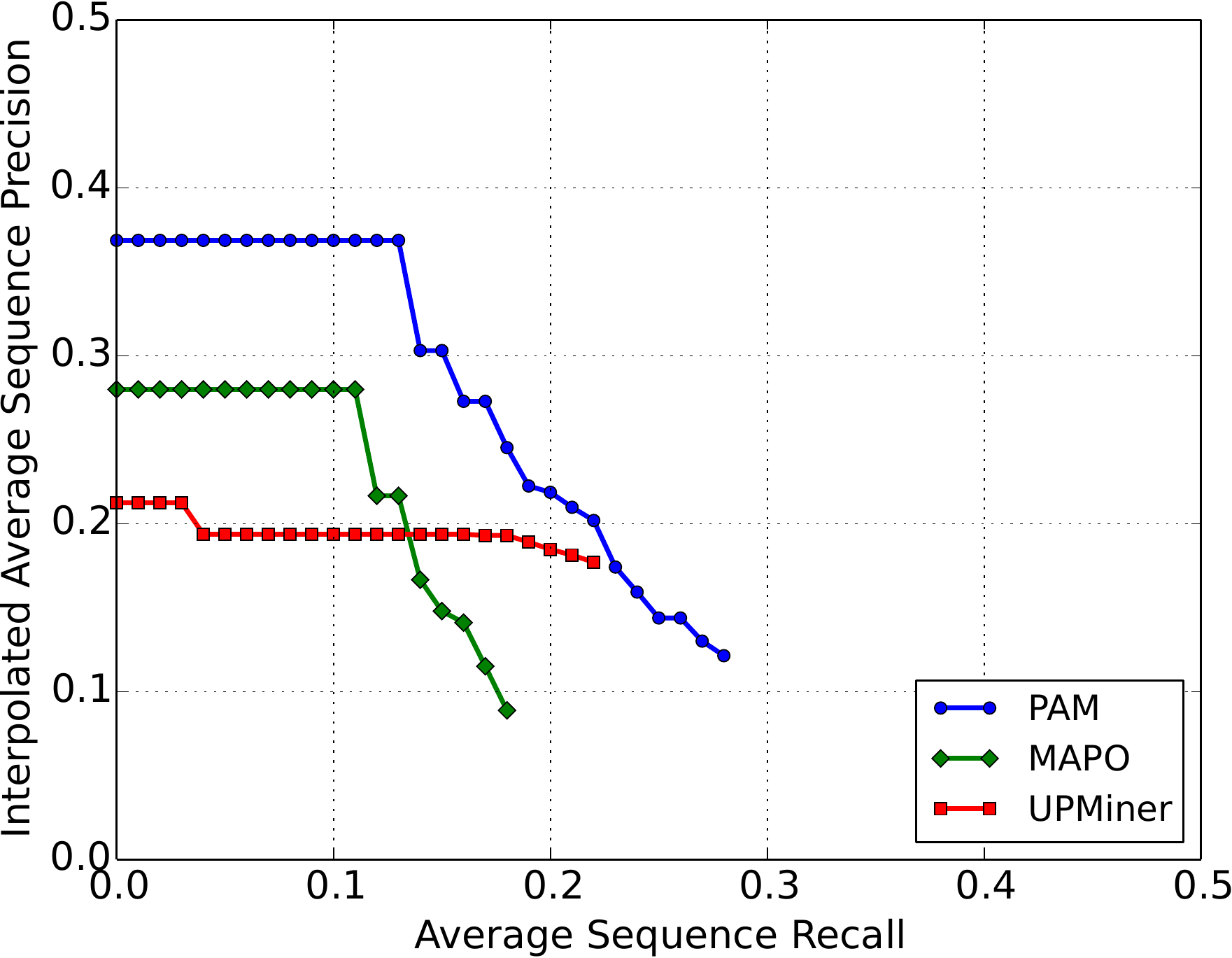}
 \caption{Average \emph{example-set} interpolated precision against recall for PAM, MAPO and UPMiner on the \textsc{Example} dataset\protect\footnotemark[2], using the top-$k$ mined sequences as a threshold.}
\label{fig:all}
\vspace{-11pt}
\end{figure}

\footnotetext[2]{This figure excludes \texttt{hadoop} as we had problems with our implementation of MAPO and UPMiner running out of memory (\texttt{hadoop} has around 2 million client LOC). While PAM had no issues, we excluded it for a fair comparison.}

\begin{figure*}[!t]
\vspace*{2pt}
\centering
\fontsize{8pt}{8pt}\selectfont\ttfamily
\begin{tabular}{lll}
TwitterFactory.<init>                         & TwitterFactory.<init>           & TwitterFactory.<init>                           \\
TwitterFactory.getInstance                    & TwitterFactory.getInstance      & TwitterFactory.getInstance                      \\
                                              &                                 &                                                 \\
Status.getUser                                & TwitterFactory.getInstance      & TwitterFactory.<init>                           \\
Status.getText                                & Twitter.setOAuthConsumer        & TwitterFactory.getInstance                      \\
                                              &                                 & Twitter.setOAuthConsumer                        \\ 
ConfigurationBuilder.<init>                   & TwitterFactory.<init>           & Twitter.setOAuthAccessToken                     \\
ConfigurationBuilder.build                    & TwitterFactory.getInstance      &                                                 \\  
                                              & Twitter.setOAuthConsumer        & Status.getUser                                  \\
ConfigurationBuilder.<init>                   &                                 & Status.getText                                  \\
TwitterFactory.<init>                         & Status.getUser                  &                                                 \\   
                                              & Status.getText                  & auth.AccessToken.getToken                       \\
ConfigurationBuilder.<init>                   &                                 & auth.AccessToken.getTokenSecret                 \\
ConfigurationBuilder.setOAuthConsumerKey      & Twitter.setOAuthConsumer        &                                                 \\
                                              & Twitter.setOAuthAccessToken     & ConfigurationBuilder.<init>                     \\
ConfigurationBuilder.build                    &                                 & ConfigurationBuilder.build                      \\
TwitterFactory.<init>                         & TwitterFactory.<init>           & TwitterFactory.<init>                           \\
                                              & TwitterFactory.getInstance      & TwitterFactory.getInstance                      \\                                                                
ConfigurationBuilder.<init>                   & Twitter.setOAuthAccessToken     &                                                 \\
ConfigurationBuilder.build                    &                                 & Status.getId                                    \\
TwitterFactory.<init>                         & ConfigurationBuilder.<init>     & Status.getId                                    \\
                                              & TwitterFactory.<init>           &                                                 \\  
ConfigurationBuilder.<init>                   &                                 & ConfigurationBuilder.<init>                     \\
ConfigurationBuilder.setOAuthConsumerKey      & ConfigurationBuilder.<init>     & ConfigurationBuilder.setOAuthConsumerKey        \\
ConfigurationBuilder.build                    & TwitterFactory.<init>           & ConfigurationBuilder.setOAuthConsumerSecret     \\  
                                              & TwitterFactory.getInstance      & ConfigurationBuilder.build                      \\
ConfigurationBuilder.setOAuthConsumerKey      &                                 & TwitterFactory.<init>                           \\
ConfigurationBuilder.build                    & auth.AccessToken.<init>         & TwitterFactory.getInstance                      \\  
                                              & Twitter.setOAuthAccessToken     &                                                 \\                     
User.getId                                    &                                 & http.AccessToken.getToken                       \\
User.getId                                    & TwitterFactory.<init>           & http.AccessToken.getTokenSecret                 \\ 
                                              & TwitterFactory.getInstance      &                                                 \\ 
                                              & Twitter.setOAuthConsumer        & Twitter.getOAuthAccessToken                     \\
                                              & Twitter.setOAuthAccessToken     & auth.AccessToken.getToken                       \\
                                              &                                 & auth.AccessToken.getTokenSecret                 \\
                                              &                                 &                                                 \\
                                              &                                 & ConfigurationBuilder.setOAuthAccessToken        \\
                                              &                                 & ConfigurationBuilder.setOAuthAccessTokenSecret
\end{tabular}
\caption{Top \texttt{twitter4j.*} API patterns mined by MAPO \cite{zhong2009mapo} (left), UPMiner \cite{wang2013mining} (middle), and PAM (right).}
\label{fig:pam_calls} 
\end{figure*}

We can however make an initial assessment of this question by going back to the held-out test set of client code that we
used in RQ1. 
We can measure how well the client test set and the handwritten examples agree
by measuring sequence precision and recall if we take the handwritten
examples as if they were API patterns and the client test set as the gold 
standard. When we do this, we find that the handwritten examples
have a recall of $27\%$, meaning that three-quarters of client API method sequences
are not contained within any of the handwritten examples. Turning to precision, the handwritten examples have a precision of $36\%$, meaning that two-thirds of API sequences from the example code
are not used by any client method (where ``used by'' means ``fully contained by'').
This is significantly lower
than the precision between the training set of client methods and the test set,
suggesting that the training set is more representative of the test set
than the handwritten examples are.
Although this might be seen as a suggestive result, we caution that
this has an important threat to validity:
handwritten examples may include scaffolding code that is unnecessary in client methods. For this reason, we advise caution about drawing strong conclusions
from the precision of handwritten examples, but we note that this threat does not apply 
to the recall. 
  
These results suggest that even in very well documented projects with extensive sets
 of examples, the API usage examples written by developers are still incomplete. 
 While it may not seem surprising that developer-written example directories would be incomplete, 
recall that we specifically
chose our data set to consist only of popular libraries with extensive 
handwritten examples --- indeed, our data set averages 18,000 lines of 
example code \emph{per target API}. 
It is striking that even with projects that are so extensively
documented, PAM is still able to infer a list of coverage
with substantially greater coverage of the API.

To gain further insight into this issue,
 we randomly selected three projects from our dataset and looked at the top five API patterns returned by PAM that were not present in any example call sequence.
We found that the 15 selected API patterns fell into the following three categories: 
7 referred to an API method not in any of the examples, 3 referred to an API class not in any of the examples and 5 referred to an API pattern that was not contained in any API example (although its methods were present in the examples).
This provides some support for the hypothesis that the API patterns document part of the API
that are used in client code but for which the original developers have not chosen to write 
specific examples.
  

Overall these results suggest that the patterns
returned by PAM could serve as a useful supplement
to code examples written by API developers. 
Indeed, these results raise the question of whether,
in future work, PAM could be used to help
 detect novel and undocumented API usages and feed them back to library and framework maintainers.

\boldpara{Qualitative Evaluation}
To provide further support to RQ3, whether the mined patterns from PAM could be useful, we
qualitatively compare and contrast the top sequences returned by PAM, MAPO, and UPMiner on an example target API. Figure \ref{fig:pam_calls} shows the top ten  mined API patterns from \texttt{twitter4j} returned by PAM, MAPO and UPMiner on the \textsc{Example} dataset.
One can clearly see that the API calls found by MAPO are extremely repetitive, in fact most of the top ten calls are just combinations of subsequences of the 
following pattern:
\begin{lstlisting}
ConfigurationBuilder.<init>
ConfigurationBuilder.setOAuthConsumerKey
ConfigurationBuilder.setOAuthConsumerSecret
ConfigurationBuilder.build
TwitterFactory.<init>
TwitterFactory.getInstance
\end{lstlisting}
which (according to our manual inspection) occurs commonly in client code but does not appear anywhere
in the top ten patterns returned by MAPO.
Similarly, the majority of the top ten UPMiner patterns are combinations of subsequences of the pattern:
\begin{lstlisting}
TwitterFactory.<init>
TwitterFactory.getInstance
Twitter.setOAuthConsumer
Twitter.setOAuthAccessToken 
\end{lstlisting}
despite the full version of this sequence appearing as the 10th pattern returned by UPMiner.
PAM on the other hand, retrieves both of these full patterns within the top ten. One might think that the \texttt{ConfigurationBuilder} pattern without OAuth returned by PAM is redundant, however not all clients use OAuth. Moreover, the sequences returned by PAM clearly
display a more diverse selection of API methods: The top ten PAM sequences
use 20 unique API methods, compared to only 8 for both MAPO and UPMiner.


\section{Conclusions}
We presented a parameter-free probabilistic API mining algorithm that makes use of a novel probabilistic model
to infers the most interesting API call patterns and demonstrated the efficacy of our approach on dataset of several hundred thousand 
API client files from GitHub. Through our experiments we found suggestions that API calls are not well documented in example code 
and in future we would like to verify this through a large-scale empirical study.

\section{Acknowledgments} 
This work was supported by the Engineering and Physical Sciences Research 
 Council (grant number EP/K024043/1).
 
\renewcommand\bibsection{\section{REFERENCES}}
\renewcommand*{\bibfont}{\raggedright}
\renewcommand{\bibsep}{5pt}
\bibliography{fse2016}
\bibliographystyle{abbrvnat}

\end{document}